\documentclass[]{aastex631}

\newcommand{\blue}{}
\let\tablenum\relax
\usepackage{siunitx}
\newcommand{\as}[1]{\ang[angle-symbol-over-decimal]{;;#1}}
\usepackage{makecell}
\usepackage{xspace} 
\newcommand{\ma}{$M_{\Gamma}$\xspace}
\newcommand{\mb}{$M_{\Gamma,\delta}$\xspace}
\newcommand{\mc}{$M_{\Gamma,\Gamma}$\xspace}
\newcommand{\ascale}{$a_{scale}$\xspace}

\newcommand{\editnon}{}
\usepackage{amsmath}

\accepted{in PSJ}

\shortauthors{Camarca et al.}

\graphicspath{{./}{figures/}}

\begin{document}
\title{A multi-frequency global view of Callisto's thermal properties from ALMA}

\correspondingauthor{Maria Camarca}
\email{mcamarca@caltech.edu}
\author[0000-0003-3887-4080]{Maria Camarca}
\affiliation{Division of Geological and Planetary Sciences, California Institute of Technology, 1200 E California Blvd M/C 150-21 Pasadena, CA 91125, USA}

\author[0000-0002-9068-3428]{Katherine de Kleer}
\affiliation{Division of Geological and Planetary Sciences, California Institute of Technology, 1200 E California Blvd M/C 150-21 Pasadena, CA 91125, USA}

\author[0000-0002-5344-820X]{Bryan Butler}
\affiliation{National Radio Astronomy Observatory, Socorro, NM 87801, USA}

\author[0000-0002-8178-1042]{Alexander Thelen}
\affiliation{Division of Geological and Planetary Sciences, California Institute of Technology, 1200 E California Blvd M/C 150-21 Pasadena, CA 91125, USA}

\author[0009-0006-2781-3484]{Cole Meyer}
\affiliation{Lunar and Planetary Laboratory, University of Arizona, Tucson, AZ 85721, USA}

\author[0000-0001-8379-1909]{Alex B. Akins}
\affiliation{Jet Propulsion Laboratory, California Institute of Technology, Pasadena, CA 91011, USA}

\author[0000-0002-4278-3168]{Imke de Pater}
\affiliation{Department of Astronomy, UC Berkeley, Berkeley, CA 94720, USA}

\author[0000-0003-0685-3621]{Mark A. Gurwell}
\affiliation{Center for Astrophysics | Harvard \& Smithsonian, 60 Garden Street, Cambridge, MA 02138, USA}



\begin{abstract}
We present thermal observations of Callisto’s leading and trailing hemispheres obtained using the Atacama Large Millimeter/submillimeter Array (ALMA) at 0.87 mm (343 GHz), 1.3 mm (233 GHz), and 3 mm (97 GHz). The angular resolution achieved for these observations ranged from \as{0.09}-\as{0.24}, corresponding to $\sim$420-1100 km at Callisto. Global surface properties were derived from the observations using a thermophysical model \citep{de_kleer_ganymedes_2021} constrained by spacecraft data. We find that Callisto's millimeter emissivities are high, with representative values of 0.85-0.97, compared to 0.75-0.85 for Europa and Ganymede at these wavelengths. \editnon{It is clear that} models parameterized by a single thermal inertia are not sufficient to model Callisto's thermal emission, and clearly deviate from the temperature distributions in the data in systematic ways. Rather, more complex models that adopt either two thermal inertia components or \editnon{that treat} electrical skin depth \editnon{as a free parameter} fit the data more accurately than single thermal inertia models. Residuals from the \editnon{global} best-fit models reveal thermal anomalies; in particular, brightness temperatures that are locally 3-5 K colder than surrounding terrain are associated with impact craters. We identify the Valhalla impact basin and a suite of large craters, including Lofn, as key cold anomalies ($\sim3-5$ K) and geologic features of interest in these data. These data provide context for Callisto JWST results \citep{cartwright_revealing_2024} as well as the other ALMA Galilean moon observations \citep{de_kleer_ganymedes_2021,trumbo_alma_2018,trumbo_alma_2017,thelen_subsurface_2024}, and may be useful ground-based context for upcoming Galilean satellite missions (JUICE, Europa Clipper).

\end{abstract}


\section{Introduction} Among the solar system's rich portfolio of icy moons, Callisto stands \editnon{out} by token of its remarkably timeworn and ancient surface. With terrains as old as \textgreater4 billion years \citep{zahnle_cratering_1998}, Callisto reigns as the geologically quiescent endmember of the  Galilean satellites. Like the other icy Galilean moons, Callisto \editnon{nominally} bears a liquid water subsurface ocean \citep{zimmer_subsurface_2000,khurana_induced_1998,kivelson_europa_1999,saur_induced_2010}, but unlike its icy siblings, Callisto does not appear to host any crustal recycling mechanisms. At present, there is little to no evidence for any volcanism, tectonics, or \editnon{other surface-interior interactions} \citep{moore_callisto_2004,greeley_galileo_2000}. Instead, Callisto's geology is dominated by a collection of craters and their evolved ruins. \editnon{At the global scale, large multi-ring impact basins such as Valhalla ($D\sim$3800 km) and Asgard ($D\sim$1400 km) located on the leading hemisphere are the dominant geologic units on this $D\sim$4820 km  moon}. At the finer scales, high-resolution images obtained by the Galileo spacecraft revealed that a dark material blankets the surface \citep{moore_callisto_2004}. Explanations for the origin of Callisto's dark material include dust infall from the irregular Jovian satellites \citep[e.g.,   ][]{bottke_black_2013}, reminiscent of how Iapetus obtained its dark leading hemisphere, and sublimation/erosional processes creating a lag deposit \citep{moore_callisto_2004}. However, the balance between exogenic and endogenic origins for this material remains poorly understood. As one of the solar system's best long-term records of impact bombardment and its largest geologically inactive moon, Callisto is the ideal target for understanding near purely exogenic surface sculpting. 

The passive thermal emission from a surface provides information on surface properties, and hence how a planetary object has evolved through interactions with its external environment. For icy, airless satellites such as Callisto, the distribution of surface heat emission is primarily controlled by material properties such as the thermal inertia $\Gamma$ (units of $\text{J}\:\text{ m}^{-2}\:\text{ K}^{-1}\:\text{ s}^{-1/2}$ are \blue{abbreviated as MKS hereafter}) and emissivity $\epsilon$ \citep{ferrari_thermal_2018}. Pinpointing how these material properties vary spatially across a surface, along with the magnitude of their variation, can help discriminate between the effect of endogenic and exogenic surface processes. For example, thermal emission measurements of the Saturnian moons Mimas and Tethys acquired with \blue{the} Cassini Composite Infrared Spectrometer (CIRS) demonstrated there is high thermal-inertia material concentrated at the apex of their leading hemispheres, which suggests that the Saturnian particle environment is capable of changing the grain properties of these moons, for example through sintering \citep{howett_high-amplitude_2011,howett_maps_2019,howett_pacman_2012}. In the Jovian system at radio wavelengths, observations \editnon{of Europa} by \cite{trumbo_alma_2017,trumbo_alma_2018} using the Atacama Large Millimeter/submillimeter Array (ALMA) showed that a purported endogenic hot spot \citep{sparks_active_2017} invoked to explain Galileo PPR data \citep{spencer_temperatures_1999} could be straightforwardly interpreted as a region of local high thermal inertia. These examples emphasize the usefulness of global or near global scale maps of planetary thermal emission. Depending on the observation wavelength used, material properties can be sensed in the uppermost surface via infrared measurements, or in the near-subsurface layers via millimeter/submillimeter measurements. A benefit of probing subsurface layers is that such observations bypass the most heavily processed surface layers. 

Space-based thermal infrared observations of Callisto come from the Voyager Infrared Interferometer Spectrometer and Radiometer (IRIS) and Galileo PPR instruments. Based on eclipse cooling curves, \cite{spencer_surfaces_1987} suggested that Callisto’s surface is not thermally uniform, with the best-fit model invoking a $\Gamma$ = 15 \blue{MKS} component overlaying a $\Gamma$ = 300 \blue{MKS} component. At a wavelength of $\sim$20 $\mu$m, the \editnon{disk-averaged} temperature was measured to be 158 K \citep{morrison_temperatures_1972,moore_callisto_2004}, while at $\sim$10 µm temperatures of $\sim$140-150 K were reported by \cite{de_pater_sofia_2021}. A measurement of the H\textsubscript{2}O ice surface temperature \editnon{derived from} spectral features is somewhat lower at 115 K, indicating that the high-albedo ice-rich regions are colder than the average surface \citep{grundy_near-infrared_1999}. Across the radio wavelength regime, disk-averaged brightness temperatures range from $\sim$135 K in the submillimeter \citep{de_pater_planetary_1989} down to $\sim$90-100 K in the centimeter \citep{pauliny-toth_brightness_1974,muhleman_precise_1986,butler_alma_2012,berge_callisto_1975, de_pater_vla_1984}, consistent with decreasing temperature with depth over the upper $\sim$meters. 

At present, global, spatially-resolved millimeter/submillimeter thermal property maps only exist for Ganymede \citep[e.g.,][]{de_kleer_ganymedes_2021} and Europa \citep{trumbo_alma_2017,trumbo_alma_2018,thelen_subsurface_2024}. Recently, \cite{camarca_thermal_2023} published a spatially resolved 343 GHz thermal image of Callisto's leading hemisphere as observed with ALMA. In that image, the Valhalla impact basin can be identified as terrain $\sim$3 K colder than model predictions, and was accompanied by low-latitude warm regions possibly linked to meteorite bombardment. However, like with the other icy Galilean satellites, a global map acquired at multiple wavelengths \editnon{is needed to identify the extent to which leading/trailing hemisphere differences exist in the millimeter, to determine whether thermal anomalies are tethered to local geologic terrains, and how the thermal emission profiles of these terrains varies with depth.}

In this work, we present the first global set of high-resolution thermal observations of Callisto using ALMA by mapping the leading and trailing hemispheres at 3 mm (97 GHz), 1.3 mm (233 GHz), and 0.87 mm (343 GHz). Section \ref{methods} describes the observations, data analysis, and flux density calibration. Section \ref{thermal-model} details the thermophysical model used to interpret the data and describes how surface properties are derived. Section \ref{results} presents the results and interpretation of the thermophysical modeling analysis. Section \ref{conclusion} summarizes the conclusions of this work.

\section{Methods} \label{methods}
\subsection{Observations}

Observations of Callisto were obtained using ALMA, located on the Chajnantor plateau in the Atacama desert of northern Chile. The main ALMA array is \editnon{composed} of 50 12-m \editnon{radio/sub-mm} antennas linked via a correlator to function as an interferometer, where each pair of antennas samples the Fourier transform of the sky brightness distribution. The data product delivered by ALMA is a measurement set (MS) \editnon{comprising} ``complex visibilities'', or the amplitudes and phases of the cross-correlated signals generated by each pair of antennas, which can be used to reconstruct the sky brightness distribution using image \editnon{inversion and deconvolution} techniques \citep[see][and references therein]{thompson_interferometry_2017}. Since the main array antennas can be distributed across up to 16 km of the plateau, ALMA is capable of achieving much higher angular resolution than is possible using a single antenna at similar wavelengths. The resolutions achieved by ALMA, in combination with its high sensitivity and spectral coverage, enable thermal mapping of the Galilean satellites and other small bodies in the solar system.

We present seven observations of Callisto's leading and trailing hemispheres obtained between 2016 October 24 and 2017 October 3 (Program 2016.1.00691.S, P.I. de Kleer). Each hemisphere has spectral coverage in three receiver bands with central frequencies of 97 GHz ($\sim$3 mm; Band 3), 233 GHz ($\sim$1.3 mm; Band 6), and 343 GHz ($\sim$0.87 mm; Band 7). The Band 6 trailing hemisphere was first observed on 2016 October 25, however the synthesized beam ($\as{0.21}\times\as{0.29}$) did not meet the resolution requirements. A corrected observation of Callisto was obtained at a similar viewing geometry on 2017 July 4 with a satisfactory synthesized beam (\as{0.12}$\times$\as{0.20}). Given that the spatial resolution of the first observation is relatively poor, we present the calibrated image and disk-averaged quantities, but do not include it in the thermal model analysis. The angular resolution of the remaining six observations ranges from $\sim$\as{0.09}-\as{0.24}. At the distance of the satellite, these angular resolutions correspond to spatial resolutions of $\sim$420 to 730 km according to the beam minor axis and scales of $\sim$700 to 1100 km on the beam major axis, suitable for mapping Callisto's largest impact basins and large-scale terrains. Within each receiver band, spectral coverage spanned four spectral windows with collective bandwidth of $\sim$8 GHz and on-source integration times of $\sim$121 s for Bands 6 and 7 to $\sim$300 s for the Band 3 observations. Calibration for array pointing, flux density, bandpass response, and phase were obtained via observations of quasars J1256–0547, J1232–0224, J1336-0829, and J1517-2422.

\begin{deluxetable*}{cccccccccccccc}
\tablenum{1}
\tabletypesize{\scriptsize} 
\tablecaption{Observing Parameters and Derived Quantities\label{table:obs-tb-table}}
\tablewidth{0pt}
\thead{\\ \\}
\tablehead{
\colhead{\thead{ \\Date\\(UT)}} & \colhead{\thead{\\Time\\(UT)}} &
\colhead{\thead{Ang.\\diam\\($\prime\prime$)}} & \colhead{\thead{\\Beam\\($\prime\prime$)}} &
\colhead{\thead{\\Beam P.A.\\($^{\circ}$)}} &
\colhead{\thead{Obs.\\ sub-Lon\\($^{\circ}$ W)}} & \colhead{\thead{Obs.\\ sub-Lat\\($^{\circ}$ N)}} &
\colhead{\thead{Phase\\Ang.\\($^{\circ}$)}} & \colhead{\thead{\\ $\nu$ \\(GHz)}} &
\colhead{\thead{\\ $\lambda$ \\(mm)}} & 
\colhead{\thead{\\$F_\nu$\\(Jy)}} &
\colhead{\thead{\\$T_b$\\(K)}} &
\colhead{\thead{Image\\rms\\(mJy $\text{beam}^{-1}$)}}&
\colhead{\thead{Image\\rms\\(K)}}
}
\startdata
2016 Nov 01 & 15:07 & 1.05 & 0.16 $\times$ 0.19 & -68.6 & 50 & -2.2 & 4.9 & 343 & 0.87 & 8.03 $\pm$ 0.40 & 116 $\pm$ 5 & 0.585 & 0.20 \\
2016 Oct 24 & 11:26 & 1.05 & 0.11 $\times$ 0.24 & 66.7 & 236 & -2.2 & 3.9 & 343 & 0.87 & 7.43 $\pm$ 0.37 & 109 $\pm$ 5 & 0.532 & 0.21 \\
2017 Jul 09 & 00:41 & 1.23 & 0.10 $\times$ 0.23 & -63.1 & 27 & -2.5 & 10.7 & 233 & 1.3 & 4.81 $\pm$ 0.24 &  109 $\pm$ 5 & 0.406 & 0.41 \\
2017 Jul 04 & 20:49 & 1.24 & 0.12 $\times$ 0.20 & -83.2 & 297 & -2.5 & 10.7 & 233 & 1.3 & 5.24 $\pm$ 0.26 & 115 $\pm$ 6 & 0.575 & 0.53 \\
2016 Oct 25 * & 12:07 & 1.04 & 0.21 $\times$ 0.29 & 60.8 & 258 & -2.2 & 4.0 & 233 & 1.3 & 3.38 $\pm$ 0.17 & 107 $\pm$ 5 & 0.610 & 0.22 \\
2017 Oct 03 & 19:33 & 1.04 & 0.09 $\times$ 0.15 & 69.5 & 84 & -2.6 & 3.2 & 97 & 3 & 0.62 $\pm$ 0.03 & 110 $\pm$ 5 & 0.193 & 1.7 \\
2017 Sep 06 & 15:34 & 1.07 & 0.11 $\times$ 0.23 & 62.0 & 224 & -2.5 & 6.7 & 97 & 3 & 0.64 $\pm$ 0.03 & 106 $\pm$ 5 & 0.184 & 0.96 \\
\enddata

\tablecomments{The $F_{\nu}$ and $T_b$ columns refer to the disk-integrated flux density and the disk-averaged brightness temperature. The Beam P.A. column refers to the beam position angle. The date marked with an * denotes a Band 6 trailing hemisphere observation acquired with a synthesized beam size larger than the proposal requirements. We opt to present the calibrated image for this poor resolution band 6 observation, but do not include it in the thermal modeling. }
\end{deluxetable*}

\subsection{Data Analysis}
To process the MS, we followed established routines for \editnon{calibrating and imaging} solar system interferometric data (e.g., see \citealt{thelen_subsurface_2024,camarca_thermal_2023, de_kleer_ganymedes_2021}), summarized below.  For a review of the interferometric observation and imaging of Solar System objects, see \cite{taylor_solar_1999}. 

\begin{figure}
\centering
\includegraphics[width=0.48\textwidth]{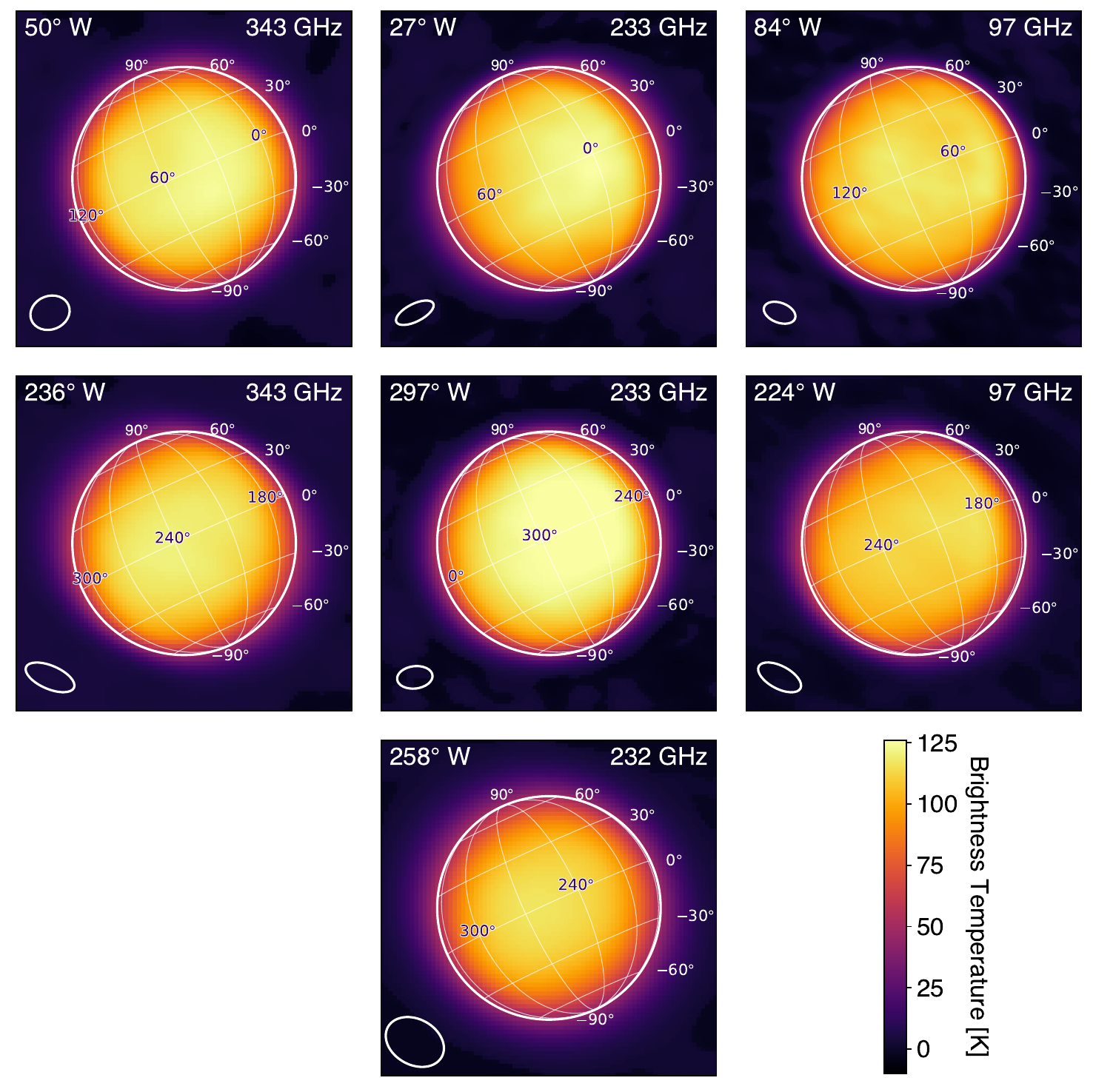}
\hfill
\includegraphics[width=0.49\textwidth]{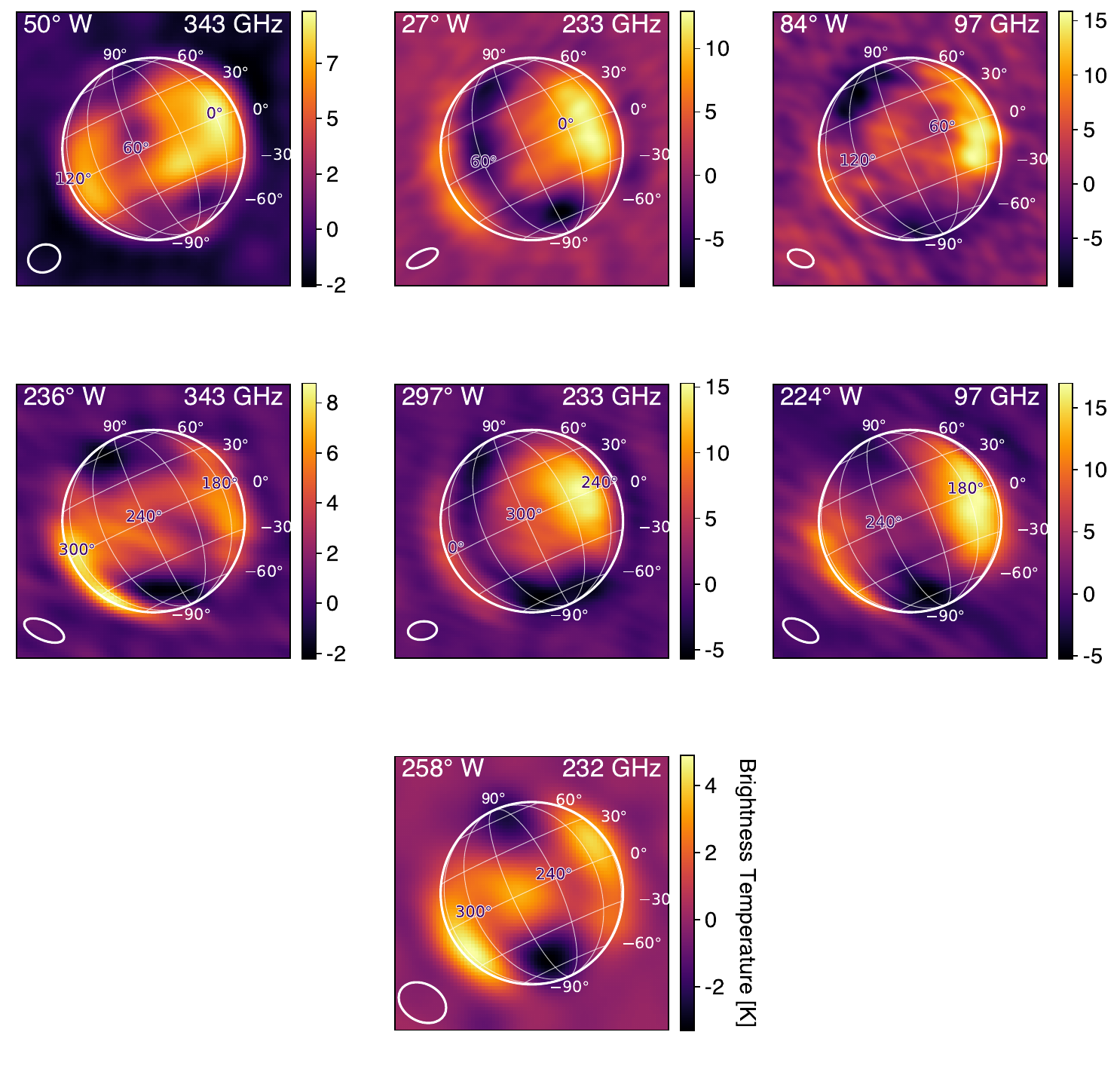}
\caption{\emph{Left}: Calibrated ALMA images of Callisto. \emph{Right:} Residuals obtained by subtracting Lambertian disks from images on the left to highlight differences in $T_b$ across the disk, \editnon{with the unit being K for all panels.} In both panels, the ellipse in the lower left corner represents the FWHM (full-width at half-maximum) of the synthesized ALMA beam, which is the resolution element. The latitude/longitude grid is spaced at 30$^{\circ}$ increments. The images are all scaled such that the horizontal and vertical axes each span -3500 km to +3500 km on a side.}
\label{fig:data}
\end{figure}

The imaging routine for these datasets consisted of an iterative cycle of imaging and self-calibration. The use of self-calibration to address phase fluctuations typically yields higher signal-to-noise images for Solar System objects than those generated directly by the ALMA pipeline. Self-calibration is effective for objects like the Galilean satellites because they have easily modeled shapes (disks) and have sufficient flux density (high SNR). For a more detailed explanation of this process, see \cite{brogan_advanced_2018}. We conducted our imaging using the Common Astronomy Software Application (CASA) package \editnon{\citep{the_casa_team_casa_2022}}. In each band, we integrated the signal over the entire 8 GHz of bandwidth using multi-frequency synthesis \citep{sault_multi-frequency_1994}.
Our phase-only self-calibration using CASA followed as such: for the first round of imaging, a Lambertian disk scaled to the size and brightness temperature of Callisto was used as a \texttt{startmodel} for the \texttt{tclean} task \citep{rau_multi-scale_2011}; no additional clean components were added. Next, the complex antenna-based gain corrections were calculated using the CASA \texttt{gaincal} task on an interval spanning the full integration time and then applied using \texttt{applycal}. Then, the corrected visibilities were imaged again adopting a shallow clean, using the previous \texttt{tclean} output model as the \texttt{startmodel}. This cycle of cleaning and computing gains continued down to an interval of 2 s; at shorter time intervals the image SNR ceased to substantially improve. For \texttt{tclean} parameters, we adopted a Briggs weighting scheme \citep{briggs_high_1995} with a robust parameter of 0.0, and used the ``Clark" deconvolver \citep{clark_efficient_1980}. The dimensions of the final clean beams are included in Table~\ref{table:obs-tb-table}. The rms (root-mean-square) noise of the final images (measured from a non-source region of the image products) is reported, with values ranging from about 0.20-1.7 K. Our final calibrated images are presented in Fig.~\ref{fig:data}.

After self-calibrating the data, we derived the disk-integrated flux density  $F_\nu$ \editnon{using a fit to the visibilities}. We report $F_\nu$ quantities from these fits rather than from the images because the visibilities are the more direct data product. Reporting $F_\nu$ is useful for placing our results in context with previous radio observations of Callisto that are largely spatially unresolved. The visibilities were fit with Bessel functions using the CASA task \texttt{uvmodelfit} assuming a uniform disk model, fitting to all spectral windows, and by excluding baselines $>$200 m for an optimal fit, as the longer baselines are sensitive to the variations across the disk rather than just the disk-integrated flux density. The final disk-averaged brightness temperature ($T_b$) was calculated by inverting the following equation (see Appendix A in \citealt{de_pater_neptunes_2014}):

\begin{equation} F{_\nu} = 10^{26}\frac{\pi R^2_C}{206265^{2}}\frac{2h\nu^3}{c^2}\times \left[\frac{1}{e^{h\nu/k_bT_b}-1} -\frac{1}{e^{h\nu/k_bT_{cmb}}-1} \right] \end{equation}
where $F{_\nu}$ is the disk-integrated flux density in Janskys, $R_C$ is the radius of Callisto in arcseconds, \editnon{the quantity 206265 is the number of arcseconds in a radian,} and, in SI units, $h$ is Planck's constant, $c$ is the speed of light, $\nu$ is the observation frequency, $k_b$ is the Boltzmann constant, and $T_{cmb}$ is the cosmic microwave background (2.7 K). The final $T_b$ measurements are presented in Table~\ref{table:obs-tb-table}. The uncertainties in the final $T_b$ incorporate flux density scale calibration errors, which \editnon{were} parameterized by a 5\% error of the total flux density, and the visibility fitting errors.
\subsection{ALMA Pipeline Flux Density Calibration Check} 
The ALMA pipeline uses quasars to calibrate the flux density of the science target data. Given that quasars are variable sources, it is necessary to verify that the ALMA pipeline has used the most complete information for setting the visibility amplitude scale. Sometimes, a manual correction to the flux density calibration scale is required (e.g, \citealt{trumbo_alma_2018,de_kleer_ganymedes_2021,thelen_subsurface_2024,de_pater_first_2019,francis_accuracy_2020}). For all of our observations, the quasar J1256-0547 was used for flux density scale calibration. To verify its flux density, measurements of the quasar taken between 2016 September and 2017 November at ALMA Bands 3, 6, and 7 were downloaded from the ALMA calibrator catalogue. These time-resolved measurements were compared to the \texttt{setjy} visibility amplitude scale set by the pipeline scripts for each day of our observations. Our inspection of a year-long range of J1256-0547 observations demonstrated consistency with the values adopted by the pipeline, and we did not make any manual corrections to the flux density scale calibration.

\section{Thermophysical Model} 
We model the data using a thermophysical model previously used for ALMA observations of the Galilean satellites, which is described in detail in Section 3 of \cite{de_kleer_ganymedes_2021}, with key points reiterated by \cite{camarca_thermal_2023} and \cite{thelen_subsurface_2024}. In this work, we summarize the main points. 

The output of self-calibrating and imaging the ALMA visibilities is a map of flux density per beam that can be converted to brightness temperature across Callisto's disk (Fig.~\ref{fig:data}). Because these observations were acquired at radio wavelengths and are therefore sensitive to subsurface emission, a thermophysical model that incorporates radiative transfer is needed \citep{mitchell_microwave_1994}. The model described by \cite{de_kleer_ganymedes_2021} treats the transport of heat via conduction and radiation through the subsurface. For each latitude and longitude on Callisto, the model constructs a time-dependent and depth-dependent temperature profile. This temperature profile is prescribed by solving the 1D heat equation, given by: \begin{equation} \rho c{_p} \frac{\partial T}{\partial t} = \frac{\partial}{\partial z}\left(k \frac{\partial T}{\partial z}\right) \end{equation} where $t$ is time, $z$ is depth, $\rho$ is the density, $c{_p}$ is the heat capacity, $T$ is the temperature, and $k$ is the thermal conductivity. For solving this differential equation, the lower boundary condition is set by the assumption of zero heat flow at depth, and the upper boundary condition is set by solar insolation at the distance of Callisto based on spatially varying \editnon{bond} albedo (we use the same spacecraft-data constrained albedo map published by \citealt{camarca_thermal_2023}). The model was evolved over $\sim$15 Callisto days (1 Callisto day = 16.69 Earth days) in time steps of at least 1/500 of a day to allow for equilibration, and was run from the surface down to to several thermal skin depths below the surface. The model is then compared to the data to determine the best-fit \editnon{global} thermal inertia and emissivity. \editnon{The reason these fits are global, rather than local, is because an individual ALMA observation is a ``snapshot" in time of Callisto's thermal emission profile. Our observations are spatially resolved, meaning we obtain the temperature at many locations across Callisto. But for an individual location, we only retrieve that temperature for a single local time. As such, we refrain from attempting to fit local properties on Callisto because we do not observe individual locations at multiple times of day. However, under the assumption that the thermal properties are constant across the disk, we can use the fact that the local time at each individual location is different to build a global thermal emission model. In other words, we
interpret spatial differences in the thermal emission as time-of-day differences. } \editnon{Additionally, we note that our modeling approach differs from \cite{de_kleer_ganymedes_2021} in that the two free parameters in our model are the thermal inertia and emissivity. In \cite{de_kleer_surface_2021}, the porosity is the primary free parameter, from which both thermophysical and electrical properties are calculated self-consistently assuming a grain size and dust fraction. In our implementation, the electrical properties are instead set assuming pure water ice, but are allowed to vary via a free scale factor. Similarly, in our implementation the thermal inertia is permitted to vary independently of the electrical properties, and without any assumptions about the underlying porosity, grain size, or composition. A final difference is that we fit the model parameters of each frequency and hemisphere independently rather than jointly fitting all data within a given frequency or longitude as done in \cite{de_kleer_ganymedes_2021}. The motivation for the different implementation was that an approach in which all data within each frequency were jointly fit with a model that requires parameters to be physically self-consistent found a poor fit to the Callisto data. We therefore opted to fit each dataset independently and increased the number of independently-tunable parameters in order to attempt to fit the data. The fact that the self-consistent model implementation was not able to fit the data indicates that some physics may be missing in the model that is relevant to Callisto but not Ganymede or Europa.}   

As described by \cite{camarca_thermal_2023}, Callisto's thermal emission at millimeter wavelengths is difficult to fit with a model that assumes a uniform thermal inertia. Therefore, after testing single-$\Gamma$ fits, we turned to testing two-$\Gamma$ models using a procedure identical to that described in Section 3.1 of \cite{camarca_thermal_2023}. This approach is an expanded application of the \cite{de_kleer_ganymedes_2021} model. In summary, we implemented the two-$\Gamma$ approach by generating a grid of ``lower" thermal inertia models ($\Gamma$ = 15-400 \blue{MKS}) and ``higher" thermal inertia models ($\Gamma$ = 500-2000 \blue{MKS}). Pairs of low-$\Gamma$ and high-$\Gamma$ models are \editnon{linearly added} at variable relative percentages prior to convolution with the ALMA beam. \editnon{At the point that the models are added, they are in units of Jy, not K}. This task produced models that simulate materials at two thermal inertias mixed spatially below the resolution of the data. Each final two-$\Gamma$ mixture model is then compared to the data. \editnon{The exploration of multiple thermal inertias for Callisto's (sub)surface emission in this work as well as by \cite{camarca_thermal_2023} is motivated in part by prior infrared results from \cite{spencer_surfaces_1987}, which showed that a single thermal inertia model is not sufficient to reproduce Callisto's infrared emission. Importantly, the \cite{spencer_surfaces_1987} treatment uses two vertically stratified layers, whereas our model treatment more closely matches a spatially inhomogeneous model.} 

A feature of the current work is an exploration of the effect of varying the electrical skin depth on \editnon{the best-fit models to} Callisto. The electrical skin depth ($\delta_{elec}$) is the material property that sets the sensitivity of an observation of wavelength $\lambda$ to thermal emission at depth $z$ below the surface. It is defined as: \begin{equation} \delta_{elec} = \frac{\lambda}{4 \pi \kappa} \end{equation}
where $\lambda$ is the observation wavelength, and $\kappa$ is the imaginary part of the complex refractive index, $\widetilde{\eta} = \eta + i\kappa$. For a detailed description of how $\delta_{elec}$ is implemented in the model, we refer the reader to Section 3.2 of \cite{de_kleer_ganymedes_2021}.  In our model implementation, $\delta_{elec}$ is computed using a $\kappa$ value that has temperature and frequency dependence appropriate for pure water ice. However, for this work we tested varying $\delta_{elec}$ by an absorptivity scaling factor \ascale that is constant in wavelength and in temperature:  \begin{equation} \delta_{elec} = \frac{\lambda}{4 \pi (a_{scale} \kappa)} \end{equation}
where the tested values of \ascale ranged from 0.25 to 12.5, \editnon{and represent the factor by which $\delta_{elec}$ is decreased. This range is informed by values obtained with similar tests of scaling the $\delta_{elec}$ for interpreting Europa ALMA data performed by \cite{thelen_subsurface_2024}}. 

The goodness of fit for the thermophysical models was determined using a cost function. \editnon{The adopted cost function} differs from a simple $\chi^2$ through an adjustment that takes into account the fact that the number of independent datapoints is smaller than the number of pixels by an amount equal to the number of pixels per ALMA beam. \editnon{This approach for estimating goodness-of-fit has been used in prior analyses, including \cite{de_kleer_surface_2021,cambioni_constraining_2019}. For each model and observation, we calculated the following:} 

\editnon{\begin{equation} \label{eq:cutoff} \chi^{2} = \frac{1}{N_{pxl}-N_{param}} \times \sum_{j=1}^{N^{j}_{pxl}} (\frac{M_{j}-D_{j}}{\sigma})^{2} 
\end{equation}}

where $N_{pxl}$ is the number of pixels on Callisto's disk, $N_{par}$ is the number of model parameters (e.g., 2 for a single $\Gamma$ and $e$), 
\editnon{Models that satisfied the relation $\chi^2<\chi^2_{min}$(1 +   \(\sqrt{2/(N_{dat}-N_{par})}\), where $N_{dat}$ is the number of individual data points, were considered acceptable \citep{press_numerical_1986,hanus_thermophysical_2015}. For our analysis, we treat $N_{dat}$ as the number of ALMA resolution elements on-disk in a given observation, rather than the number of individual pixels on-disk; this results in a more conservative estimate of model uncertainties. To calculate $N_{dat}$, we divide the total number of pixels on-disk in a given observation by the number of pixels per ALMA beam.} For additional details on applying the cost function to ALMA data, see Section 3.4 of \cite{de_kleer_surface_2021}.

\editnon{For ease of readability, we implement a naming scheme into this work. The thermal model approaches used herein will be referred to with a capital M, with subscripts denoting the key variables of interest. $M_{\Gamma}$ will refer to the single-$\Gamma$ treatment, $M_{\Gamma,\Gamma}$ to the two-$\Gamma$ treatment, and $M_{\Gamma,\delta}$ to the treatment using a single $\Gamma$ and variable $\delta_{elec}$.}  
 
\label{thermal-model}
\section{Results \& Discussion} \label{results}
We present surface properties inferred from ALMA images of Callisto obtained at ALMA Bands 3, 6, and 7, with corresponding wavelengths/frequencies of 3 mm (97 GHz), 1 mm (233 GHz), and 0.87 mm (343 GHz), respectively. These observations provide coverage of both Callisto's leading and trailing hemispheres. Along with the calibrated images shown in Fig.~\ref{fig:data}, we present best-fit thermophysical properties, as well as the residual maps produced by subtracting the corresponding best-fit models from the data \editnon{to identify spatial variations in thermal properties}. In Section \ref{integrated-brightness}, we present disk-averaged $T_b$ measurements, and place our results in context with past thermal wavelength observations. In Sections \ref{emiss} and \ref{global}, we present the best-fit thermal properties and describe global trends in the spatial distribution of thermal emission.  Lastly, in Section \ref{local-residuals} we detail the connections between regional thermal signatures and the local geology.

\subsection{Disk-Averaged Brightness Temperature}\label{integrated-brightness} 
\begin{figure}
\centering
\includegraphics[scale=0.43]{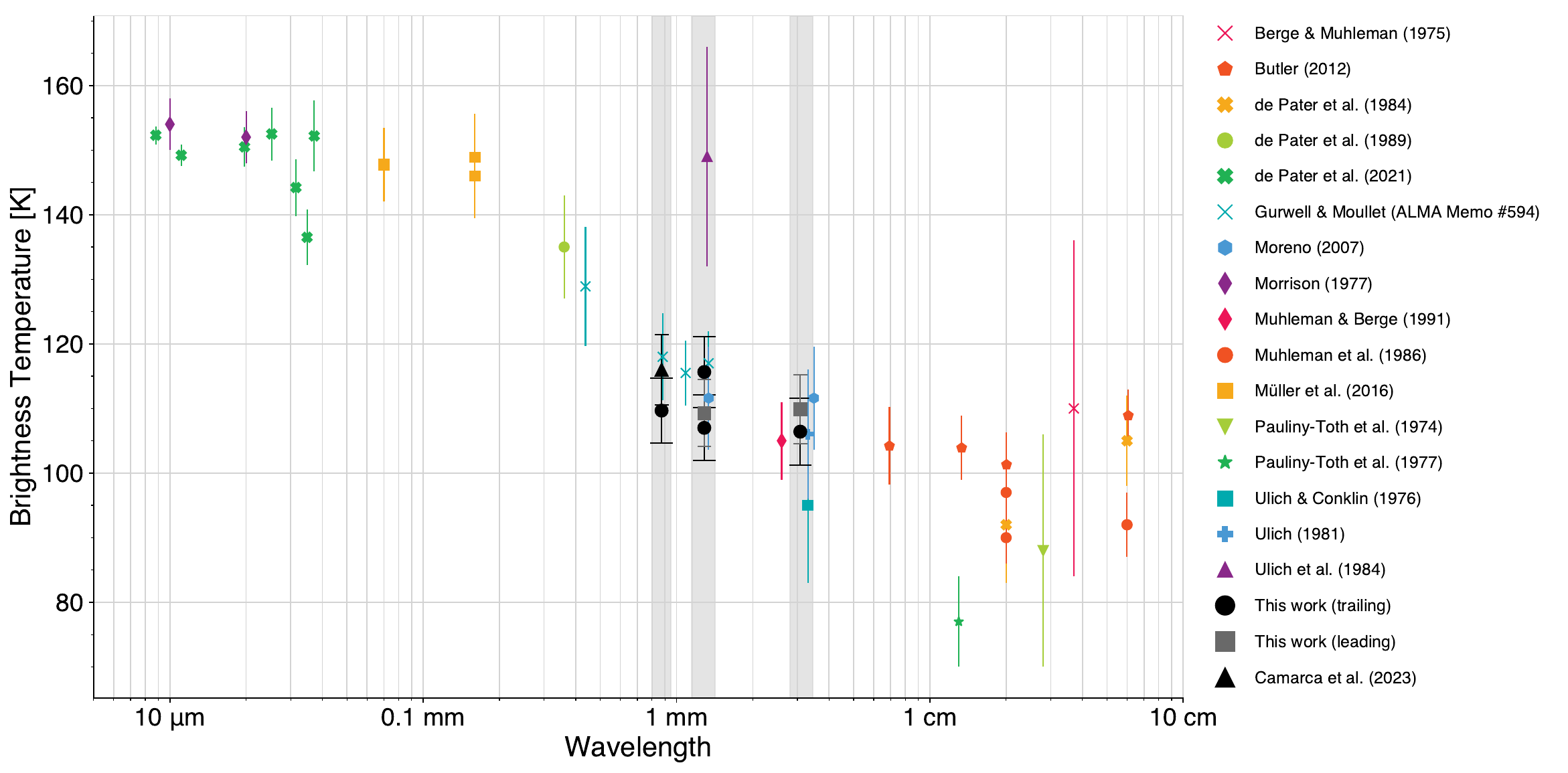}
\caption{Summary of disk-\editnon{averaged} brightness temperature measurements for Callisto plotted as a function of wavelength. This plot is an update to the one shown in \cite{camarca_thermal_2023} including the additional ALMA data points presented here. Our measurements (black/gray points with capped error bars, \editnon{highlighted in grey}) agree with most neighboring data. Data in this plot are taken from: \cite{berge_callisto_1975}, \cite{butler_alma_2012}, \cite{de_pater_vla_1984,de_pater_planetary_1989,de_pater_sofia_2021}, Gurwell \& Moullet (ALMA memo \#594, \citealt{butler_alma_2012}), \cite{moreno_report_2007}, \cite{morrison_galilean_1977}, \cite{muhleman_observations_1991}, \cite{muhleman_precise_1986}, \cite{muller_far-infrared_2016}, \cite{pauliny-toth_brightness_1974,pauliny-toth_observations_1977}, \cite{ulich_observations_1976}, \cite{ulich_millimeter-wavelength_1981}, and  \cite{ulich_planetary_1984}.}
\label{fig:all-disk-integrated}
\end{figure}

Because most past radio observations of Callisto are not spatially resolved, it is useful to report disk-integrated quantities to place our results in context. In Table~\ref{table:obs-tb-table}, we report the disk-integrated flux densities $F_\nu$ in Jy and corresponding disk-averaged brightness temperatures $T_b$ in K for the leading and trailing hemispheres at each frequency. In Fig.~\ref{fig:all-disk-integrated}, we place these disk-averaged measurements in context with past measurements obtained for Callisto over the cm to sub-mm range. We find the $F_\nu$ span $\sim$0.62-8.03 Jy across the wavelength range of 3 to 0.87 mm, corresponding to $T_b$ values in the range of $\sim$106-116 K. Consistent with what was previously reported by \cite{camarca_thermal_2023} for the 0.87 mm leading hemisphere measurement, the additional ALMA measurements presented here agree well with past work. However, the 1.3 mm measurement by \cite{ulich_planetary_1984} of $T_b$\editnon{=} 149 $\pm$ 17 K is about $\sim$35 K higher than other 1.3 mm measurements, including those presented in this work as well as those by Gurwell and Moullet (personal communication) and \cite{moreno_report_2007}. The measurement of Ganymede by \cite{ulich_planetary_1984} also disagrees with other measurements: in Fig. 6 of \cite{de_kleer_ganymedes_2021}, the 1.3 mm $T_b$ reported by \cite{ulich_planetary_1984} is $\sim$115 K, while nearby ALMA measurements \citep{de_kleer_ganymedes_2021} and IRAM–PdB \citep{moreno_report_2007} fall within the 80-100 K range (for Ganymede, the error bars of \cite{ulich_planetary_1984}'s $T_b$ overlap with the neighboring measurements; for Callisto, the error is not close to overlap). 

There is a general trend of decreasing $T_b$ with increasing wavelength, beginning with values near 150 K in the infrared near 10 $\mu$m that fall off to 90-100 K in the 1-10 cm regime (Fig.~\ref{fig:all-disk-integrated}). The depression of $T_b$ values with increasing wavelength is likely due to the fact that longer wavelengths sense deeper subsurface layers where 1) the diurnal wave is attenuated and 2) higher thermal inertia (i.e., denser, more compacted) materials are expected to be present.\editnon{ In addition, there is drop in emissivity at millimeter wavelengths that is known to be linked to the optical properties of ice. At longer wavelengths, water ice becomes much more transparent than it is in the far infrared due to a decrease in the value of the imaginary part of the complex index of refraction. (e.g., \citealt{warren_optical_2019}). This behavior is also well documented in observations of the icy particles of Saturn's rings, for example, which exhibit a drop-off in emissivity around 0.2 mm \citep{spilker_cassini_2005,tiscareno_thermal_2018}}. We find the $T_b$ measurements for Callisto's leading and trailing hemisphere at a given frequency are \editnon{the same} within uncertainties, a result that differs from hemispheric disparities observed in Europa ALMA and Submillimeter Array (SMA) observations (see \citealt{thelen_subsurface_2024} and references therein).

\subsection{Emissivity}\label{emiss}
Nearly all of the thermal models that provide acceptable fits to the Callisto data have high millimeter emissivities ($\epsilon$), with values ranging from $\epsilon\sim$0.85-0.97. As shown in Figs.~\ref{fig:single-ti}, \ref{fig:ed-fits}, and \ref{fig:two-ti}, Callisto's high emissivities persist across the 97-343 GHz range, across hemispheres, and across thermophysical model treatments. Even for observations that yield unconstrained $\Gamma$ (e.g., the single-$\Gamma$ treatment of the 97 GHz data), the derived emissivities remain high and overlap with the emissivity derived from observations that produced a more well-constrained $\Gamma$, demonstrating the robustness of these emissivity values. Prior measurements of Callisto's emissivity in other wavelength regimes also returned high values. As highlighted by \cite{moore_callisto_2004}, emissivities near unity in the microwave were inferred from observations at 0.355 mm by \cite{de_pater_planetary_1989} and beyond 2 mm by \cite{muhleman_observations_1991}. In the infrared, \cite{spencer_surfaces_1987} found that Callisto's emissivity approached values up to around 0.92-0.94 in Voyager data.

These measurements of $\epsilon$ for Callisto can be placed in context with values derived for the other Galilean moons. A direct $\epsilon$ comparison is available via published ALMA observations of Ganymede and Europa. For Europa, \cite{trumbo_alma_2018} reported $\epsilon$ = 0.75 at 233 GHz based on thermal model treatment of just surface emission, while \cite{thelen_subsurface_2024} reported values of $\epsilon$ = 0.80-0.85 across frequencies of 97.5, 233, and 343.5 GHz after accounting for subsurface emission. For Ganymede, \cite{spencer_thermal_1987} found a high infrared emissivity of 0.94, while \cite{de_kleer_ganymedes_2021} reported $\epsilon$ = 0.75-0.78 in the millimeter. To summarize, Callisto's emissivities are consistently higher than its icy Galilean siblings across our frequency coverage, which is likely a consequence of the fact that Callisto's surface is not as abundantly ice-rich as its Galilean siblings and that the emissivity of ice is lower than rock at millimeter wavelengths \citep{hewison_airborne_1999,yan_retrieval_2008,de_kleer_surface_2021}.

\subsection{Global Thermal Properties}\label{global} 
We derived global thermal properties for Callisto by fitting a thermal model to the ALMA data using a suite of thermophysical modeling approaches, including single-$\Gamma$ and two-$\Gamma$ models, variations to the electrical skin depth, and the implementation of refraction using Hapke and Fresnel laws. We found most of the Callisto observations are not well-fit using models with a uniform thermal inertia across the satellite's surface, motivating the more complex approaches. The introduction of a secondary thermal inertia component and decreasing the electrical skin depth did improve the fit for a number of cases. In this section, we describe the outcomes of each individual modeling approach.

\subsubsection{Single-$\Gamma$ \editnon{R}esults}
As seen in the first row of Fig.~\ref{fig:model-resids}, the data-minus-model residual images obtained by using a single-$\Gamma$ approach (\ma) produced systematic features in the residuals from nearly every Callisto observation. For instance, the models usually failed to simultaneously fit the satellite disk center and the disk limb, with a typical model producing overly cold limbs and an overly warm disk center. The best-fit $\Gamma$ and $\epsilon$ values are shown in Fig.~\ref{fig:single-ti}. We find the emissivities inferred from \ma are consistently high across all observations, with values generally above 0.90, although the range extends from 0.86 to 1.0 depending on the frequency band (values exceeding $\epsilon$ = 1 were not permitted). Regarding the inferred $\Gamma$ values, there is a large degree of variation between frequencies. The 343 GHz best-fit $\Gamma$ ranges from about $\Gamma$ = 600-1900 \blue{MKS}, with no appreciable difference between the leading and trailing hemisphere. The 233 GHz data behave somewhat differently, with the leading hemisphere observation fit best by a much narrower range, $\Gamma$ = 100-500 \blue{MKS}, while the trailing hemisphere range of $\Gamma$ = 200-1300 \blue{MKS} is wide compared to the 343 GHz data, albeit with a lower average value. The 97 GHz data are the most poorly constrained observations, although there are two separate ranges allowed: the first being $\Gamma$ = 100-600 \blue{MKS} and the other being $\Gamma$ = 1200-2000 \blue{MKS}. 

A poorly constrained $\Gamma$ at 97 GHz is not surprising, given the 97 GHz data are sensitive to deeper subsurface layers than the 343 GHz and 233 GHz data. As described in Sec.~\ref{integrated-brightness}, deeper subsurface layers experience more muted diurnal temperature fluctuations, which makes differentiating between models more difficult. A similar ALMA Band 3 result for Europa was reported by \cite{thelen_subsurface_2024}. Despite the lack of a constrained $\Gamma$, the 97 GHz trailing hemisphere observation is the only observation for which the \ma treatment does not result in image residuals that exhibit large-scale spatial systematics indicative that the model is not a good fit to the data. Rather, the 97 GHz trailing residuals are more complex, with isolated regions of warm and cold terrains; these localized residuals will be discussed in Section \ref{local-residuals}. \blue{In this work, we use ``systematics'' to generally refer to large residuals resulting from the inability of certain models to reproduce both the center/edge emission simultaneously.  For example, the leading hemisphere 233 GHz data are poorly fit using the \ma treatment, which yields a warm disk and cold limbs. However, adding an additional component, either a secondary thermal inertia or a scalable skin depth, removes these major systematics and reveals a residual map that appears more well connected to underlying geologic features (e.g., a more obvious cold spot near Valhalla consistent with the 345 GHz model residuals). }

\subsubsection{Single-$\Gamma$ and Variable $\delta_{elec}$ \editnon{R}esults}
We explored a variant of the \ma approach where we allowed variations to \editnon{$\Gamma$ and to} the electrical skin depth $\delta_{elec}$ (\mb), which determines the subsurface depths that are responsible for the model emission at a given wavelength. The default electrical skin depth in the model corresponds to that of pure water ice. It depends on both temperature and frequency, but for the representative case of 233 GHz and 150 K, \editnon{we calculate it} is 65$\lambda$. The range of absorptivity scaling factors explored therefore corresponds to $\delta_{elec}$ in the range of $\sim$5-150$\lambda$. The low end of this range is representative of rocky regolith, while the upper end is representative of pure and porous ice \editnon{(e.g., see Fig. 2 in \citealt{de_kleer_ganymedes_2021} and also \citealt{de_kleer_surface_2021})}. As shown in Fig.~\ref{fig:ed-fits}, we find that the \editnon{model variables for the 233 GHz data are more tightly constrained than those of} the 343 or 97 GHz data, which is consistent with the model behavior found  in the single-$\Gamma$ approach. The direction of decreasing $\mathbf{\chi^2}$ for the models trends toward fractionally lower $\delta_{elec}$ values (higher values of \ascale). Nearly all of the models with the lowest individual $\mathbf{\chi^2}$ for a given observation adopted an absorptivity scaling factor of $\sim$10-12.5, which represents the upper bound of \ascale tested in the current work. Similarly, \cite{thelen_subsurface_2024} found that \editnon{the $\delta_{elec}$ required a scaling factor of order 10} to interpret Europa's thermal emission at 233 GHz.

For many models, allowing this parameter to vary provides an improvement over the \ma fits. As shown on the second row of Fig.~\ref{fig:model-resids}, several model fits are greatly improved with the \mb approach. The two observations most responsive to the \mb treatment include the leading hemisphere 233 and 97 GHz images, with more modest improvements to the remaining observations. For example, the 233 and 97 GHz model residuals under the default \ma treatment are dominated by the warm disk centers and cold limbs described in the previous subsection. With the addition of $\delta_{elec}$ as a free parameter, the residuals are no longer dominated by \editnon{large-scale spatial systematics} but rather show localized residuals that are correlated with geologic features.

\subsubsection{Two-$\Gamma$ \editnon{R}esults}
We additionally modeled all of the observations using a two-$\Gamma$ model (\mc{}), which \cite{camarca_thermal_2023} found to provide a better fit to Callisto ALMA data than \ma models. We show the best-fit \mc  residuals in the third row of Fig.~\ref{fig:model-resids}. For some observations, the inclusion of a secondary thermal inertia component provided an improved fit over \ma. For example, the disk/limb fit problems for leading hemisphere 343 and 233 GHz images are muted using \mc. In the 343 GHz data, the \mc approach greatly improved the ability of the model to match Callisto's cool limbs, especially the (geographic) eastern limb; additionally, the improved fit resulted in more prominent residual $T_b$ signatures that can be linked with geologic terrain, i.e. the Valhalla impact basin. Similarly, in the 233 GHz leading hemisphere image, a cold spot co-located with Valhalla is not visible in the \ma residuals, but is visible with the improved \mc residuals because it is no longer dominated by systematic residuals associated with a global model that is not a good match. By contrast, only modest improvement (or change) to the residuals is seen for the 97 GHz leading hemisphere data and the trailing hemisphere data across all frequency bands. For instance, the trailing hemisphere 343 GHz and 233 GHz \mc residuals still bear the center-of-disk/limb fit systematics apparent in the \ma residuals, albeit with smaller amplitude temperature deviations. The range of \mc thermal inertia combinations that satisfied Eq.~\ref{eq:cutoff} for each observation are presented in Fig.~\ref{fig:two-ti}. Generally, the $\Gamma$ constraints for the 343 GHz data are better compared to the 233 GHz data. Like with the \ma modeling prescription, the 97 GHz \editnon{models} are very poorly constrained. Additionally, the distribution of the 97 GHz trailing hemisphere temperature residuals (i.e., the relative location of the hot and cold spots) is nearly identical to the \ma model. Collectively, these trends with frequency indicate that $\Gamma$ is better constrained at higher frequencies because the diurnal temperature variations are larger due to the shallower depths sensed, while the temperature residuals observed at low frequencies are more robust to $\Gamma$. \editnon{Although there are limitations to this approach, if high/low thermal inertia terrains are adjacent to each other at distances well below the resolution of an ALMA beam (hundreds of km), this is a reasonable approximation. An example of such terrains include the swaths of bright, icy peaks interspersed between Callisto's dark material blanket seen in high-resolution spacecraft images \citep{greeley_geology_2001,moore_callisto_2004}}.  

\subsubsection{Summary of Global Results} 
We find that a thermophysical model that adopts an approach more complex than just fitting for a single $\Gamma$ and $\epsilon$ for the surface provide better fits to Callisto's thermal emission as measured at 97, 233, and 343 GHz. \editnon{When the \ma approach is used, we find the best-fit values derived from \ma as organized by frequency are $\Gamma$ = 600-1900 \blue{MKS}, 100-1300 \blue{MKS}, and unconstrained, for 343, 233, and 97 GHz respectively. However, modeling approaches that adopt either two-$\Gamma$ components or treat the electrical skin depth $\delta_{elec}$ as a free parameter generally found a better fit, and in some cases resolved persistent large-scale spatial residuals untreated by the \ma approach. Using \mb, we find $\Gamma$ = 500-2000, 300-500, and 50-2000 \blue{MKS} for 343, 233, and 97 GHz, respectively.  Application of the \mc approach produced  fits for the (lower, higher) components of $\Gamma$= (\textless 100, unconstrained) \blue{MKS} for 233 GHz; $\Gamma$ = (\textless50, \textgreater1200) \blue{MKS} for 343 GHz; and totally unconstrained for 97 GHz.} The fact that Callisto's regolith appears to have a very high emissivity, with representative values ranging from $\sim$0.85-0.97 in our frequency regime, is consistent across all model treatments. 

\begin{figure}
\centering
\includegraphics[scale=0.3]{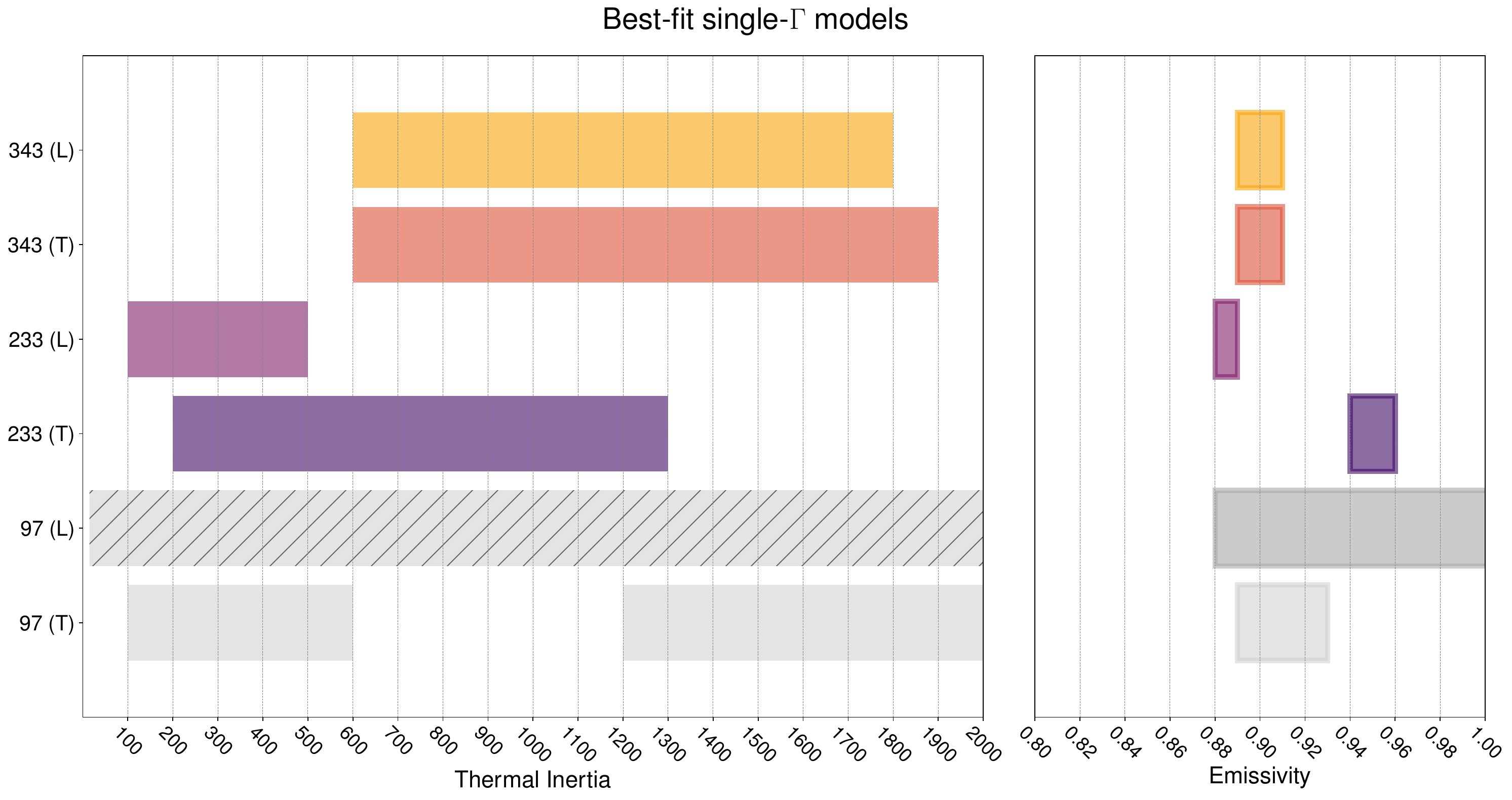}
\caption{Results from model fits using single-$\Gamma$ models (\ma). We note that the \ma approach produced consistent systematic effects in the residual maps indicating that none was a good fit, and is therefore not the preferred thermal modeling approach. For each observation frequency and hemisphere, the range of $\Gamma$ \editnon{and emissivity} values that satisfied \editnon{Eq.~\ref{eq:cutoff}} are indicated by colored bars. The 97 GHz (L) models are not constrained in $\Gamma$ space and are marked by a hatch pattern. The range of best-fit millimeter emissivities are shown on the right panel. \blue{The units of $\Gamma$ are $\text{J}\:\text{ m}^{-2}\:\text{ K}^{-1}\:\text{ s}^{-1/2}$}}
\label{fig:single-ti}
\end{figure}

\begin{figure}
\centering
\includegraphics[scale=0.3]{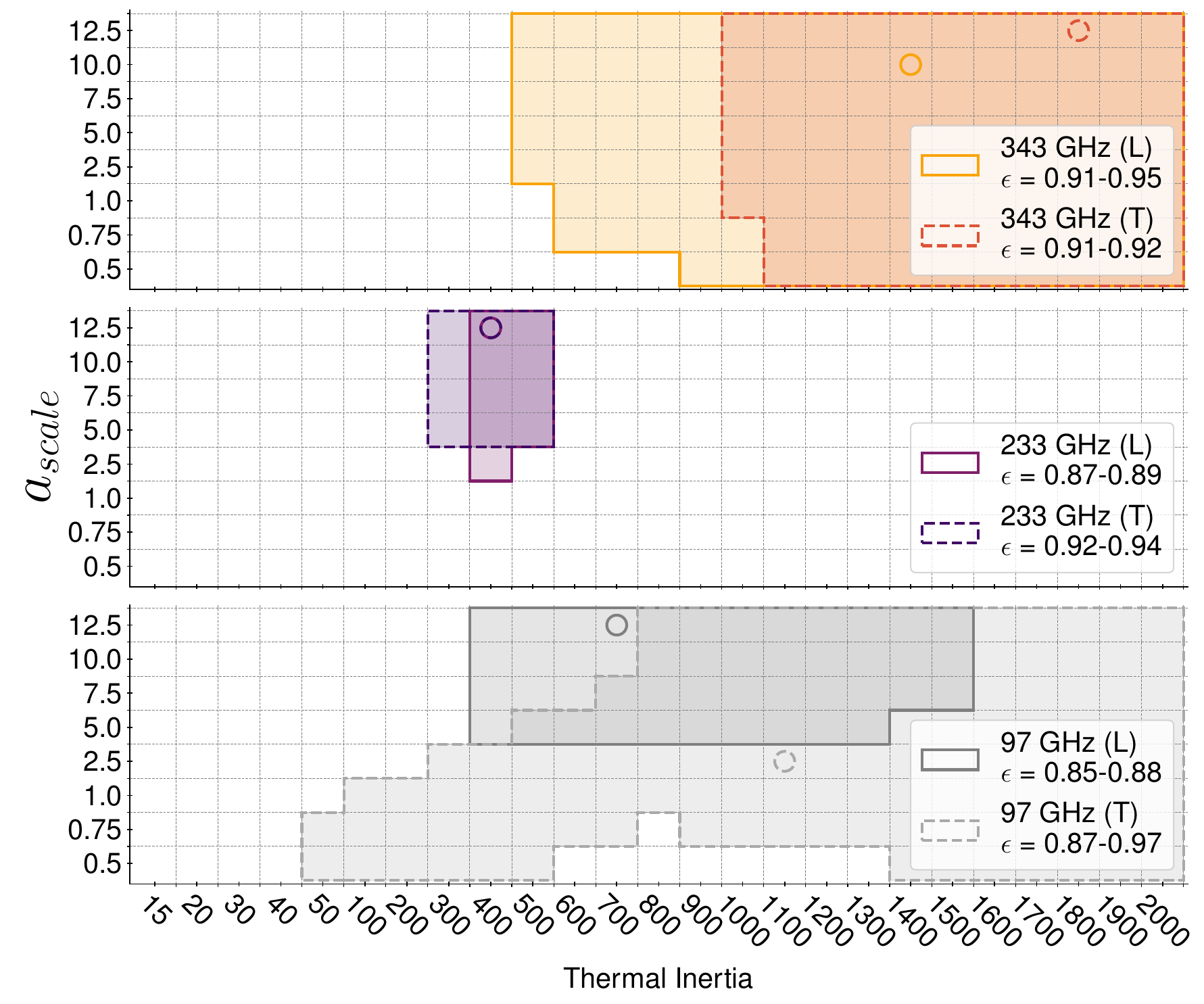}
\caption{Summary of model fits \editnon{treating} electrical skin depth \editnon{as a free parameter, \mb. The y-axis is presented as the factor by which $\delta_{elec}$ is decreased relative to pure water ice (\ascale); the actual $\delta_{elec}$ these values correspond to are temperature-dependent. Shaded regions refer to models that} satisfied \editnon{Eq.~\ref{eq:cutoff}}. The model results for each observing frequency (343, 233, and 97 GHz) are presented on separate panels for clarity.  The circular markers indicate the location of the minimum $\chi^2$ for each set of models; additionally, these markers indicate the \mb models plotted in the second row of Fig.~\ref{fig:model-resids}. Out of all of the observations, the 233 GHz observations are the most \editnon{sensitive to} changing $\delta_{elec}$, while the observations at the other frequencies yield \editnon{more poorly} constrained $\delta_{elec}$ ranges. Importantly, this figure should be interpreted alongside Fig.~\ref{fig:model-resids}, which demonstrates visually how the \mb models outperform the \ma models. \blue{The units of thermal inertia are $\text{J}\:\text{ m}^{-2}\:\text{ K}^{-1}\:\text{ s}^{-1/2}$}}
\label{fig:ed-fits}
\end{figure}

\begin{figure}
\centering
\includegraphics[scale=0.43]{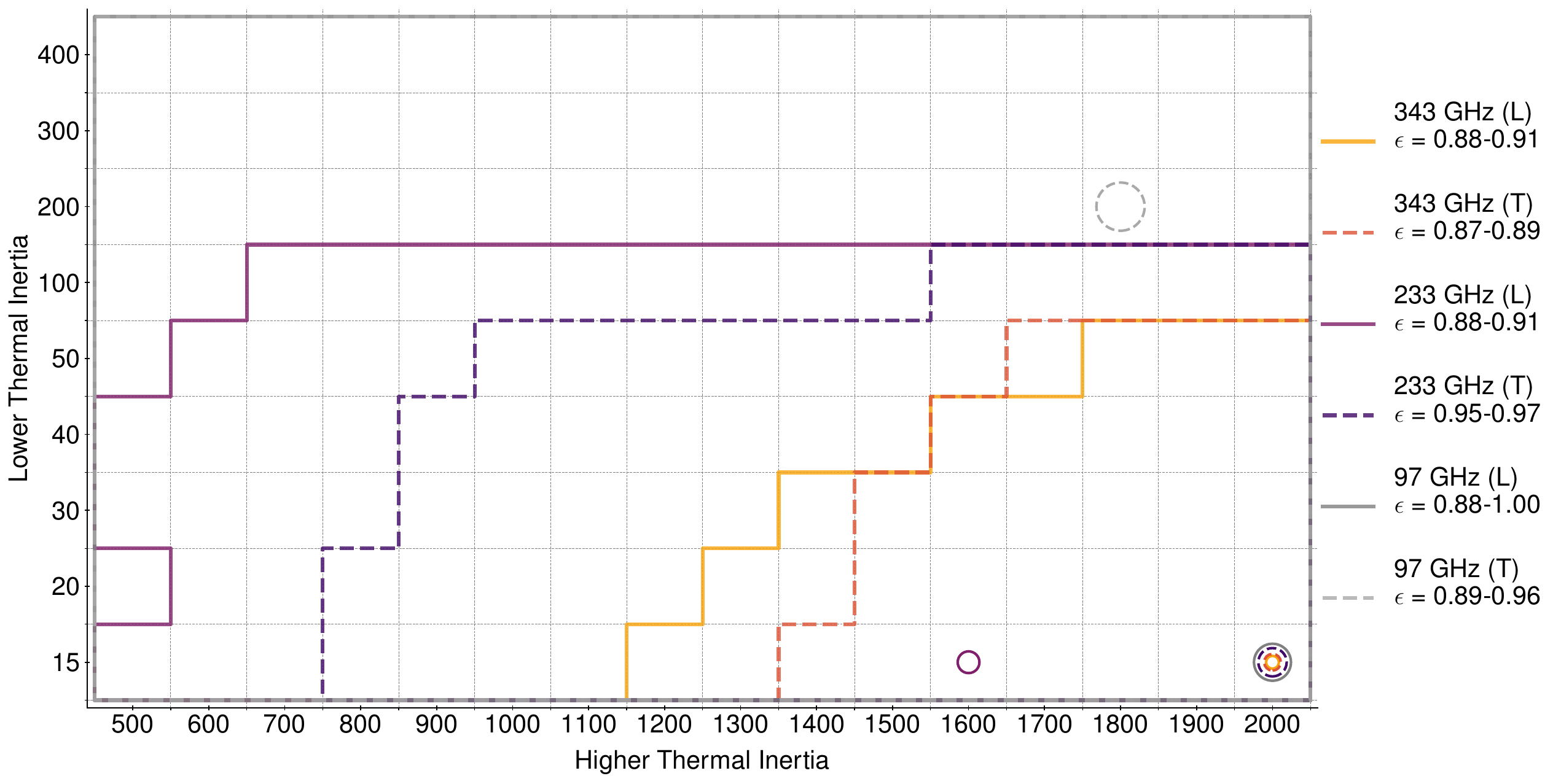}
\caption{Summary of the two-$\Gamma$ model (\mc) fits. For each observation, the range of two-$\Gamma$ mixtures that satisfied \editnon{Eq.~\ref{eq:cutoff}} are enclosed below and to the right of the labeled lines. The 97 GHz results include the entire parameter space. The best-fit mixture model for each observation is denoted by an open circle; these same models are plotted in the third row of Fig.~\ref{fig:model-resids}. The legend text (L) or (T) indicates a leading or trailing hemisphere observation, respectively. The best-fit emissivities are indicated under the observation label in the legend. For ease of interpretation, the exact ranges of acceptable mixing percentages for a given pair of thermal inertias is not shown.  Generally, the $\Gamma$ constraints for the 343 GHz data are tighter compared to the 233 GHz data. The 97 GHz data, both leading and trailing hemisphere, are not constrained at all. \blue{The units of thermal inertia are $\text{J}\:\text{ m}^{-2}\:\text{ K}^{-1}\:\text{ s}^{-1/2}$}}
\label{fig:two-ti}
\end{figure}

Regarding these $\Gamma$ values, there is not a clear trend with subsurface depth.  Both Fig.~\ref{fig:single-ti} and Fig.~\ref{fig:two-ti} demonstrate that the preferred thermal inertias appear higher for the 343 GHz data than the 233 GHz data, but there is overlap between the inferred values. \editnon{Moreover, we do not constrain a difference in the value of the thermal properties between the leading and trailing hemispheres that is consistent across all model approaches. This is apparent in  Fig.~\ref{fig:single-ti},~\ref{fig:ed-fits}, and ~\ref{fig:two-ti}, which shows that the thermal properties for a given frequency overlap between the two hemispheres. Lastly, the persistent inability of the modeling approaches to fix the large-scale warm/cold regions on the trailing hemisphere is suggestive that additional model physics may be required for those data.} \blue{The fact that we do not constrain a difference in Callisto's thermal properties between 97-343 GHz leaves open the possibility that our ALMA observations have not bypassed the dark material blanket that globally covers Callisto. Estimates of the thickness of Callisto's dark material are up to many $\sim$meters to $\sim$10s of meters thick based on surface topography, including the smoothness of the surface at the 10 meter scale \citep{moore_callisto_2004,basilevsky_morphology_2002}. Therefore, it was not expected a priori that the ALMA data would bypass the dark material blanket, and the present modeling results appear consistent with this expectation. Altogether, the ALMA data may offer a lower bound for the dark material blanket that is not less than $\sim$10s of cm thick.}
                                     
\editnon{To place these $\Gamma$ results in further context, we summarize results from ALMA observations of the other Galilean satellites. Using eclipse cooling data, \cite{de_pater_alma_2020} inferred Io's thermal inertia to be $\Gamma$ = 50 \blue{MKS} based on infrared data, and around $\Gamma$ = 350 \blue{MKS} based on millimeter observations. At 233 GHz, \cite{trumbo_alma_2018} obtained $\Gamma$ = 95 \blue{MKS} for Europa, with possible surface variations of $\Gamma$ = 40-300 \blue{MKS}. A comprehensive analysis of Europa at 97-345 GHz by \cite{thelen_subsurface_2024} retrieved effective thermal inertias of  $\Gamma$ = 50–140 \blue{MKS}, $\Gamma$ = 140–180 \blue{MKS} for the leading and trailing hemispheres, respectively. The Ganymede observations by \cite{de_kleer_ganymedes_2021} at 343.5, 223, and 97.4 GHz yielded effective thermal inertias of $\Gamma$ = 450$\substack{+300\\-250}$, 350$\substack{+350\\-250}$, and 750$\substack{+200\\-350}$ \blue{MKS}, respectively. Although the thermal modeling procedure is not the same across all of these works, these findings, together with the results presented by \cite{camarca_thermal_2023}, it appears there is some trend toward increasing thermal inertia with increasing distance from Jupiter (a trend also illustrated in Fig. 6 of \citealt{camarca_thermal_2023})}. 

\subsection{Local Thermal Residuals}
\label{local-residuals}
\begin{figure}
\centering
\includegraphics[width=1.0\textwidth]{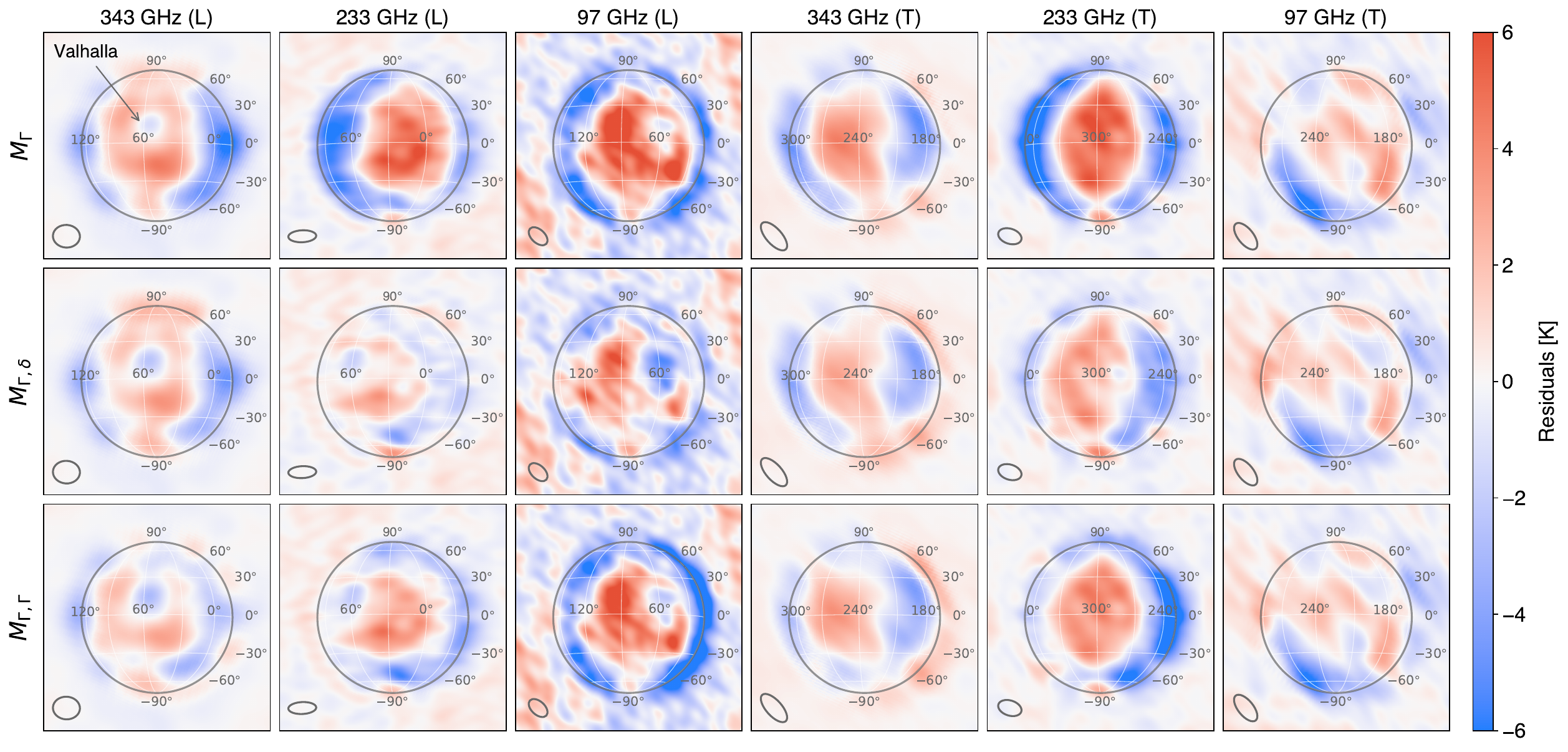}
\caption{Grid of residuals derived from different thermal modeling approaches. Each column represents a unique observation, and each row is dedicated to a different modeling treatment. The first row shows residuals derived from models generated using a best-fit single thermal inertia (\ma). The second row shows residuals using a best-fit single thermal inertia with a variable skin depth (\mb). The third row shows best-fit residuals from using the two-thermal inertia mixture approach (\mc). Individual thermal anomalies on the disk edge marked in grey are not considered reliable.}
\label{fig:model-resids}
\end{figure}

\begin{figure}
\centering
\includegraphics[width=1.0\textwidth]{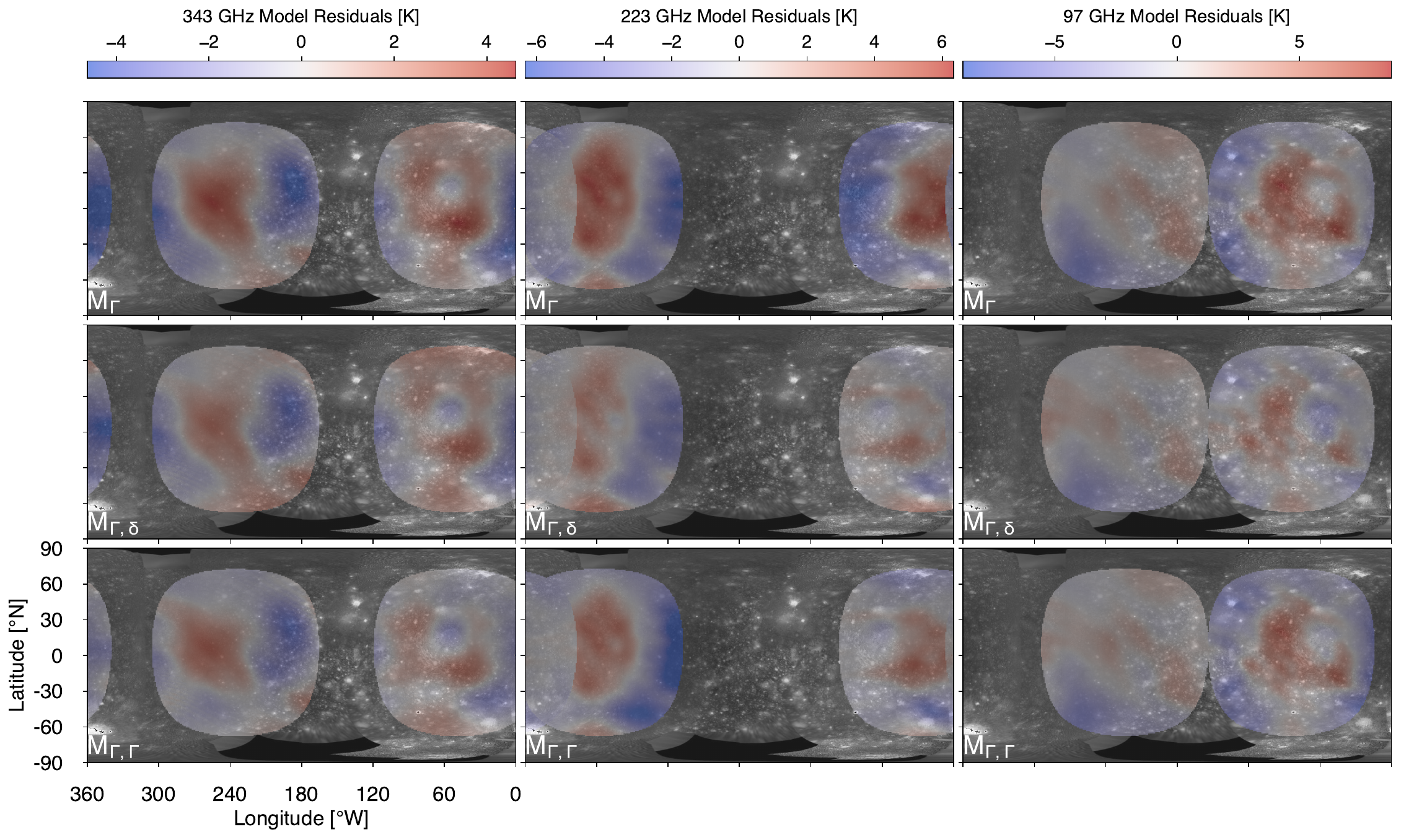}
\caption{Residuals from Fig.~\ref{fig:model-resids} shown in projected form with the grayscale USGS Callisto mosaic based on spacecraft data shown in the background \blue{(mosaic retrieved from \url{https://astrogeology.usgs.gov/search/map/callisto_galileo_voyager_global_mosaic_1km})}. The Callisto grayscale data has been slightly modified to bring out surface features. Temperature residuals from each of our three modeling efforts are overlain in color, with warm colors denoted regions where the measured data are warmer than model predictions, and blue regions are colder than model predictions. First row: $T_b$ residuals obtained using \ma models. Second row: $T_b$ residuals obtained using \mc models. Third row: $T_b$ residuals obtained using \mc models. The 343 GHz and 233 GHz trailing hemisphere model fits appears to have systematic \editnon{discrepancies from the observations} (i.e., poor center of disk/limb fits), while residuals at other hemispheres/frequencies appear to provide more reliable fits.  Individual thermal anomalies near the disk center are generally more reliable than those at the disk edge.}
\label{fig:model-proj}
\end{figure}

In the previous section, we outlined the global thermophysical properties we derived for Callisto. In this section, we describe thermal anomalies \editnon{that may be tethered to the local geology. In these regions, the data do not agree with the global model, indicative of distinct thermal properties.} For thermal anomalies that do not have an obvious geologic connection, we speculate on possible explanations within the context of other observations of Callisto. The model residuals are presented in non-projected form in Fig.~\ref{fig:model-resids}, as well as in map projected form in Fig.~\ref{fig:model-proj}.

\subsubsection{Valhalla Impact Basin}
One of the most prominent thermal anomalies in these data is a cold spot on the leading hemisphere that tracks the center of the Valhalla impact basin (Fig.~\ref{fig:model-resids},~\ref{fig:model-proj}). With its outer troughs extending to around 3800 km, Valhalla is Callisto's \editnon{largest} impact scar, \editnon{and is likely the} largest multiring impact basin in the solar system \citep{moore_callisto_2004}. The center of Valhalla is located at $\sim$18$^{\circ}$N and $\sim$57$^{\circ}$W, and was in-view for all of our leading hemisphere observations. For the 343 GHz data presented in \cite{camarca_thermal_2023}, the site of the Valhalla impact basin was associated with a thermal anomaly about 10 K colder than surroundings in the calibrated \editnon{images} (rms of 0.20 K) and 3-5 K colder than the surroundings after subtracting the best-fit global model (i.e., after accounting for albedo). In the new 233 and 97 GHz images presented here, a similar large cold spot presents itself, which can be seen in the calibrated images and is even more pronounced in the Lambertian disk subtracted images ( Fig.~\ref{fig:data}). The $T_b$ deviations of Valhalla from surrounding terrain in the 97 and 233 GHz best-fit thermal models are $\sim$5.2 K (3$\times$ rms) and  $\sim$1.8 K (4$\times$ rms), respectively. The 233 GHz data observed Valhalla close to the morning terminator, while the 97 and 343 GHz data viewed it at local afternoon. The fact that the temperature difference with surrounding terrain is so much lower at 233 GHz therefore indicates that this region \editnon{might have} a higher thermal inertia than the surrounding terrain. It is possible that a lower emissivity in this region is also contributing to the thermal anomaly.

The association of cold thermal anomalies with large impact craters has been observed across many icy solar system objects. On Ganymede, cold spots were associated with the large impact craters Tros (D$\sim$94 km) and Osiris (D$\sim$107 km), as well as clusters of unresolved, small bright craters \citep{de_kleer_ganymedes_2021}. A 3 m depth temperature map based on 0.6-22 GHz observations using the Juno MWR instrument also confirmed that Tros is cold \citep{brown_microwave_2023}. Although Europa has far fewer craters than Ganymede or Callisto, its young crater Pwyll is thermally cold, as confirmed by \cite{trumbo_alma_2017,trumbo_alma_2018} and \cite{thelen_subsurface_2024}. The Saturnian satellites follow suit, with cool craters identified on Titan \citep{janssen_titans_2016} and Rhea \citep{bonnefoy_rheas_2020}. In the cases listed here, cold is defined relative to a thermal model that already accounts for albedo. We note that, while cold craters are a general theme in thermal wavelength data, there are certainly exceptions to this trend. For example, the large crater Tashmetum on Ganymede was not substantially cold in resolved ALMA images, suggesting some threshold for crater freshness may be required to impact a thermal fingerprint at these wavelengths \citep{de_kleer_ganymedes_2021}. Importantly, the Valhalla results presented here show that the cold crater trend is a phenomenon observed even for the largest impact class among icy satellites, namely the multiring impact basin. Understanding the thermal properties of large craters on Callisto may help inform further studies of localized volatile retention/excavation. For example, \citep{cartwright_revealing_2024} recently mapped the 4.25 $\mu$m $\mathrm{CO_2}$ absorption feature on Callisto using the Near Infrared Spectrograph (NIRSpec) instrument onboard the James Webb Space Telescope (JWST). These maps revealed solid-phase $\mathrm{CO_2}$ enhancements in the multiring impact basin Asgard, as well as in pixels sampling the edge of Valhalla (Valhalla itself was not observed), a result in agreement with the Galileo NIMS map \citep{hibbitts_co2-rich_2002}. 

 \subsubsection{Adlinda, Heimdall, and Lofn}
 Like with Valhalla, the new data presented here also indicate the persistence of a cold spot near a suite of craters in Callisto's southern hemisphere (Fig.~\ref{fig:model-resids},~\ref{fig:model-proj}). As noted in \cite{camarca_thermal_2023}, a spot about 3-4 K cooler than nearby terrain was observed at \SI{40}{\degree} S, \SI{12}{\degree} W  in the 343 GHz thermal model residuals. The candidate geologic unit(s) for that cold residual is the Adlinda/Heimdall/Lofn impact complex, which was not resolved by the ALMA beam. However, the cold spot's proximity to the satellite's limb precluded a definitive assignment to a local geologic feature due to geometric foreshortening. However, the new 233 GHz leading hemisphere image, centered at a sub-observer longitude of \SI{27}{\degree} W, sampled a more favorable viewing geometry of the Adlinda/Heimdall/Lofn terrain, and indeed a cold spot at $\sim$\SI{30}{\degree} W, $\sim$\SI{45}{\degree} S and is near this leading hemisphere crater suite. \blue{In the albedo map underlay of Fig.~\ref{fig:model-proj}, Lofn is identifiable as the largest, bright feature near this lat/lon.} The sub-observer longitude of the 97 GHz leading hemisphere data was \SI{85}{\degree} W, which is too far west to observe any of these craters. 
\subsubsection{Cold Spot Near \SI{270}{\degree} W --- a $\mathrm{CO_2}$ Gas Connection?}
One of the primary thermal features in the 97 GHz trailing hemisphere image is a cold spot roughly centered at \SI{-60}{\degree} N and \SI{270}{\degree} W (Fig.~\ref{fig:model-resids}, ~\ref{fig:model-proj}). The 97 GHz trailing hemisphere data are not particularly responsive to different thermal model treatments, but depending on the implementation used, the cold spot is approximately 5-6 K colder than nearby terrain. Based on the image rms of 0.96 K, this cold region is detected at \editnon{5$\times$ rms}, making it the dominant thermal feature in that image; for comparison, the warmest warm region in that image is only $\sim$3 K above the best-fit model. Before proceeding to a discussion of this feature, we note that its proximity to the satellite's limb necessitates caution. This feature is additionally not obvious in the 233 GHz and 343 GHz trailing hemisphere image residuals. However, those data are poorly fit by the range of models tested, and it is the case that the cold spot tentatively associated with Adlinda/Heimdall/Lofn in \cite{camarca_thermal_2023} using just the 343 GHz image, which did not have a favorable viewing angle for that crater complex, is confirmed to be real based on the additional 233 GHz data presented here that has a more favorable viewing geometry. Treating this 97 GHz southern cold spot as tentatively real, we now highlight other relevant observations in this region. 

Given that the sub-observer longitude of this image is \SI{224}{\degree} W, the nominal surface coverage includes material spanning \SI{134}{\degree}- \SI{314}{\degree} W. The possibility of whether the outermost terrains of Heimdall could explain this feature was explored, but in the \cite{greeley_galileo_2000} geologic map of Callisto, crater ejecta units mapped for Heimdall span from \SI{315}{\degree}-\SI{10}{\degree} W, suggesting that these mapped materials are not relevant. \blue{Additionally, an inspection of the albedo map in Fig.~\ref{fig:model-proj} demonstrates there are no major bright impacts (i.e., multi-ring basins or other large features) in this vicinity.} Recently, \cite{cartwright_revealing_2024} published \editnon{solid-phase} $\mathrm{CO_2}$ band depths and $\mathrm{CO_2}$ gas column \editnon{density} for Callisto using the NIRSpec instrument onboard JWST. The JWST map reproduced the same trailing $\mathrm{CO_2}$ bullseye pattern originally seen in Galileo NIMS data \citep{hibbitts_co2-rich_2002}; on Callisto, the trailing hemisphere is far richer in \editnon{solid-phase} $\mathrm{CO_2}$ than the leading due to radiolytic processes. Curiously, in the JWST NIRSpec trailing hemisphere image cube, Callisto's peak $\mathrm{CO_2}$ gas column \editnon{density} is physically offset from the peak of the solid $\mathrm{CO_2}$ absorption. Instead, the gas peak is offset at around \SI{-45}{\degree} N, well aligned with where we observe the cold thermal anomaly in the ALMA data. It is possible that ALMA is sensing a local variation in subsurface composition that is linked to the CO$_2$ gas distribution.

\subsubsection{Additional local residuals}
While the previous two subsections described localized thermal anomalies for which there were geologic associations, in this section we discuss thermal features that lack such an association. There is a cold spot on the trailing hemisphere centered at roughly \SI{210}{\degree} W and \SI{-20}{\degree} N that is visible in a number of model residuals. The feature is most prominent in the 97 GHz trailing hemisphere residual map, \editnon{which does not exhibit prominent disk center/limb fit systematics}. However, evidence for this feature may also exist in the 343 GHz trailing residuals, as there is possibly a slightly colder structure at the same location in those model residuals. Additionally this feature can be seen in the 343 GHz trailing hemisphere Lambertian disk subtracted image (Fig.~\ref{fig:data}). Based on the \cite{greeley_geology_2001} geologic map, the disk in-view for the 97 GHz trailing hemisphere image does not sample any major geologic units. The terrain is largely cratered plains with smaller mapped craters that are much smaller than the ALMA synthesized beam. A mapped geologic unit in the \cite{greeley_geology_2001} map that approaches the ALMA beam size is a rounded patch of light plains centered at around 290 W and \SI{-28}{\degree} N. The light plains units are marked by a higher albedo (compared to the typical cratered plains), bear fewer craters, and are thought to form from the more ice-rich materials excavated by impact processes. Additionally in the 97 GHz trailing hemisphere residual, near \SI{30}{\degree} N and \SI{180}{\degree} W, there is a cold spot observed that is about 1.3 K colder than the disk average. This feature is possibly due the presence of some craters that fall just within the western bounds of the multiring basin Asgard. The primary named crater in this region is the bright impact feature Nirkes.

In addition, we follow up on the discussion in \cite{camarca_thermal_2023} addressing the existence of an ``L-shape" thermal feature on the leading hemisphere in the 343 GHz feature. To summarize the finding of \cite{camarca_thermal_2023}, there appears to be an L-shaped warm anomaly occupying the area west and south of Valhalla. The vertical component of the L is centered at almost \SI{90}{\degree} W and is symmetric across the equator with a divergence from the model of $\sim$2-3 K, while the horizontal portion of the L is similarly warm, and extends eastward to $\sim$\SI{25}{\degree} W. The 233 and 97 GHz leading images presented here provide more face-on viewing geometries for the horizontal and vertical components of this warm L, respectively. In Fig.~\ref{fig:model-resids},\ref{fig:model-proj}, the 233 GHz residuals also bear the warmest $T_b$ values in the area east and south of Valhalla, with a very similar morphology to that observed in the 343 GHz residuals. In the 97 GHz image, a warm region is reproduced in the same area, although the other parts of the ``L" shape appear less structured. In \cite{camarca_thermal_2023}, micrometeorite bombardment preferentially texturizing Callisto's leading hemisphere (and therefore resulting in lower $\Gamma$ materials that can warm up faster) was proposed as a possible origin of this leading hemisphere warm anomaly. However, it is not clear from the model residuals whether or not Callisto's leading hemisphere is significantly warmer than the trailing. The 97 GHz residuals do seem to indicate that leading hemisphere low-latitude warm spots are warmer than trailing hemisphere low-latitude warm spots, however the 233 and 343 GHz trailing hemisphere residuals have systematic patterns that preclude a similar inspection. \editnon{The large scale structure of the hot and cold spots on the trailing hemisphere as observed in the 233 and 343 GHz residuals may require more complex model physics (e.g., roughness) than is used by the present approach.} Future efforts to model Callisto's unique thermal properties may address these issues. 

\section{Conclusions} \label{conclusion}
We present millimeter thermal images of Callisto's leading and trailing hemisphere obtained at 97 GHz (3 mm), 233 GHz (1.3 mm), and 343 GHz (0.87 mm). Our observations were acquired using the Atacama Large Millimeter/submillimeter Array. We find \editnon{disk-averaged} brightness temperatures across the 97-343 GHz regime to be $T_{b}$ = 106-116, and are consistent with past (unresolved) observations. 

To interpret our data, we used three different thermal modeling procedures: 1) \ma: single-$\Gamma$ models, 2) \mb: single-$\Gamma$ models with a variable electrical skin depth $\delta_{elec}$, and 3) \mc: two-$\Gamma$ models.  Consistent with the conclusions of \cite{spencer_thermal_1987} and \cite{camarca_thermal_2023}, Callisto's thermal emission is not well modeled if only a simple single-$\Gamma$ treatment is used. In this work, we find the \ma approach yielded systematic residuals across the disk, including a persistent inability to fit the disk limb and disk center simultaneously. We find that, for several observations, the addition of a \editnon{variable} $\delta_{elec}$ as a free parameter or a secondary $\Gamma$ yielded a better fit to the data, as evidenced by an increased ability to fit both the disk center and disk limb emission, as well as lower overall $T_b$ residuals. For example, a thermal cold spot associated with the Valhalla impact basin is barely discernible in \ma 233 GHz leading hemisphere residuals, but is easily discernible in the \mb and \mc treatments. \editnon{A robust finding is consistently high millimeter emissivities across all model treatments, ranging from 0.85-0.97, compared to 0.75-0.85 for Europa \citep{trumbo_alma_2017,trumbo_alma_2018,thelen_subsurface_2024} and Ganymede \citep{de_kleer_ganymedes_2021} at these wavelengths}. \editnon{To summarize the thermal inertias derived from each of the three model approaches (\ma,\mb,\mc) we find: the best-fit values derived from \ma are $\Gamma$ = 600-1900, 100-1300 \blue{MKS}, and unconstrained, for 343, 233, and 97 GHz respectively; we highlight that we find the \ma approach is not preferred for these Callisto data. The best-fit values adopting the \mb approach are: $\Gamma$ = 500-2000, 300-500, and 50-2000 \blue{MKS} for 343, 233, and 97 GHz respectively. Lastly, the best-fit (lower, higher) $\Gamma$ ranges from \mc are: $\Gamma$ = (\textless 100, unconstrained) \blue{MKS} for 233 GHz; $\Gamma$ = (\textless50, \textgreater1200) \blue{MKS} for 343 GHz; and unconstrained for 97 GHz.} \blue{The fact that we do not constrain an obvious difference in the value of the thermal properties across the sampled 97-343 GHz frequency range may suggest that these ALMA data, which probe depths of up to $\sim$10s of cm, have not bypassed Callisto's dark material blanket. }

Callisto's trailing hemisphere observations were generally much less responsive to thermophysical model alterations compared to the leading hemisphere. The 343 GHz and 233 GHz trailing hemisphere residuals both bear systematic $T_b$ east/west structure that is not accounted for within our current model parameters. Because of this, identifying localized thermal features for those observations is difficult. Although the $\Gamma$ values are also difficult to constrain for the 97 GHz trailing hemisphere data, regardless of model treatment, the residuals do not have the same systematic structure that persists at the other two higher frequencies.

After subtraction of the best-fit thermal models, the residual images reveal several thermal anomalies. There are 3-5 K cold thermal anomalies associated with Valhalla, as well as with the impact craters Adlinda/Heimdall/Lofn (which are not resolved from one another in the data). \editnon{Regarding Valhalla, its thermal anomaly may be explained by the presence of a material that is either higher thermal inertia than surrounding terrain, and perhaps lower emissivity}. An additional cold anomaly is seen in the trailing hemisphere southern latitudes, which is co-located with the maximum CO$_2$ gas observed by JWST/NIRSpec \citep{cartwright_revealing_2024}. Finally, an L-shaped warm anomaly in the center of the leading hemisphere tentatively presented by \cite{camarca_thermal_2023} in the standalone 343 GHz data is confirmed at additional frequencies, and may be due to a lower thermal inertia regolith there due to micrometeorite bombardment. 

The data presented here provide the first global, spatially-resolved thermal maps of Callisto’s surface at ALMA observing frequencies, and collectively shed light on the material properties of Callisto's bulk surface as well as providing a new window into the surface materials present in its large impact basins. \blue{In the near-future, Callisto's thermal emission may be revisited after the ALMA wideband sensitivity upgrade, which will permit this level of $T_b$ precision at much higher spatial resolution without requiring prohibitively long exposure times. Additionally, these ALMA data may be combined with Very Large Array (VLA) observations of Callisto \citep{akins_resolved_2024}, which samples the 1-50 GHz regime, as well as observations taken with the planned next generation VLA (ngVLA). The lower frequencies accessible by these observing facilities could reach \textgreater meter depths, and could help provide further constraints on Callisto's dark material blanket. Moreover, future spacecraft observations of Callisto by the JUICE instrument payload will provide ground-truth data to complement observations from telescope facilities. The JUICE instrumentation particularly relevant to these ALMA data includes the Submillimeter Wave Instrument (SWI), which operates at 530-1275 GHz, and will help link existing millimeter and infrared observations. Altogether, a new era of Callisto exploration is in the near future, and continued studies of Callisto's thermal properties will shed light on the origins of the surface material of this ancient moon.}


\section*{Acknowledgements}
We acknowledge support from the National Science Foundation through a Graduate Research Fellowship under Grant No. DGE‐1745301 to M.C., as well as through grants 2308280 and 2308281, which supported K.d.K., M.C., A.E.T., and I.d.P. This research was also funded in part by the Heising-Simons Foundation through grant \#2019-1611. Funding was additionally provided by the NASA ROSES Solar System Observations program (through Task Order 80NM0018F0612) for A.E.T., K.d.K., A.A. Contributions by A.A. were carried out at the Jet Propulsion Laboratory, California Institute of Technology, under a contract with the National Aeronautics and Space Administration (80NM0018D0004). This paper makes use of the following ALMA data: ADS/ JAO.ALMA\#2016.1.00691.S. ALMA is a partnership of ESO (representing its member states), NSF (USA) and NINS (Japan), together with NRC (Canada), MOST and ASIAA (Taiwan), and KASI (Republic of Korea), in cooperation with the Republic of Chile. The Joint ALMA Observatory is operated by ESO, AUI/NRAO, and NAOJ. The National Radio Astronomy Observatory is a facility of the National Science Foundation operated under cooperative agreement by Associated Universities, Inc.
\bibliography{refs.bib}
\bibliographystyle{aasjournal}
\end{document}